\documentclass[12pt,preprint]{aastex}
\def\1{\'\i}
\shorttitle{The TSZ signal of WMAP 3yr data.}
\received{December 3, 2007}
\begin{document}
\title{Measurement of the electron-pressure profile of galaxy clusters in 
Wilkinson Microwave Anisotropy Probe (WMAP) 3-year data.}
\author{F. Atrio-Barandela\altaffilmark{1,5}, A. Kashlinsky\altaffilmark{2}
D. Kocevski\altaffilmark{3}, H. Ebeling\altaffilmark{4}}
\altaffiltext{1}{F\'{\i}sica Te\'orica, Universidad de Salamanca, 37008 Salamanca, Spain}
\altaffiltext{2}{SSAI and Observational Cosmology Laboratory, Code 665,
Goddard Space Flight Center, Greenbelt MD 20771}
\altaffiltext{3}{Department of Physics, University of California at Davis, 1 Shields Avenue,
Davis, CA 95616}
\altaffiltext{4}{Institute of Astronomy, University of Hawaii, 2680 Woodlan Dr, Honolulu, HI 96822}
\altaffiltext{5}{e--mail: atrio@usal.es}

\begin{abstract}
Using WMAP 3-year data at the locations of close to $\sim 700$
X-ray selected clusters we have detected the amplitude of the
thermal Sunyaev-Zeldovich (TSZ) effect at the 15$\sigma$ level,
the highest statistical significance reported so far. Owing to the
large size of our cluster sample, we are able to detect the
corresponding CMB distortions out to large cluster-centric radii.
The region over which the TSZ signal is detected is, on average,
four times larger in radius than the X-ray emitting region,
extending to $\sim 3h_{70}^{-1}$Mpc. We show that an isothermal
$\beta$ model does not fit the electron pressure at large radii;
instead, the baryon profile is consistent with the
Navarro-Frenk-White profile, expected for dark matter in the
concordance $\Lambda$CDM model. The X-ray temperature at the
virial radius of the clusters falls by a factor $\sim 3-4$ from
the central value, depending on the cluster concentration
parameter. Our results suggest that cluster dynamics at large
radii is dominated by dark matter and is well described by
Newtonian gravity.
\end{abstract}

\keywords{Cosmic Microwave Background. Cosmology: theory. Cosmology: observations.}

\section{Introduction.}

The hot intergalactic X-ray emitting gas distorts the Cosmic
Microwave Background (CMB) spectrum of those photons crossing the cluster.
The CMB distortions are independent of 
redshift and arise from Thompson scattering; when electrons are
non-relativistic, they are caused by two different effects: (1)
the Thermal Sunyaev-Zel'dovich effect (TSZ, Sunyaev \& Zel'dovich
1972) is due to the thermal motions of electrons in the cluster
potential well, whereas (2) the Kinematic Sunyaev-Zel'dovich
effect (KSZ) is due to the motion of the cluster as a whole. Both
effects contribute to the CMB radiation power spectrum, but their
contribution is significant at $\ell>10^3$ (Atrio-Barandela \&
M\"ucket 1999, Molnar \& Birkinshaw 2000). For the most luminous
clusters, relativistic corrections could become important, giving
rise to the relativistic SZ effect (Wright, 1979). The TSZ
spectral signature appears as a temperature decrement in the WMAP
frequency range. It has been measured for $\sim 100$ individual
clusters (Birkinshaw 1998, Carlstrom, Holder \& Reese 2002,
LaRoque et al 2007), while the KSZ effect, being of much smaller
amplitude, has not yet been detected for individual systems. It
can, however, be detected statistically with large cluster samples
(Kashlinsky \& Atrio-Barandela, 2000) using a method recently
applied by us to measure the bulk motion of clusters on scales out
to $\sim 300h^{-1}$Mpc (Kashlinsky et al 2007).

Currently planned ground- and space-borne CMB experiments like the
South Pole Telescope, the Atacama Cosmology Telescope and the PLANCK
mission are expected to detect clusters via their TSZ signature in the
near future. In the meantime, first efforts to determine the TSZ
contribution to the CMB fluctuations observed by WMAP across the
entire sky (Spergel et al 2007) were made by cross-correlating
templates constructed from galaxy and cluster catalogs using 1st-year
data (Hern\'andez-Monteagudo \& Rubi\~no Mart\1n 2004,
Hern\'andez-Monteagudo, G\'enova-Santos \& Atrio-Barandela 2004, Myers
et al 2004, Afshordi, Lin \& Sanderson 2005) and, more recently,
3rd-year data (Afshordi et al 2007). These studies report 2-8$\sigma$
detections of an anti-correlation with various galaxies and cluster
surveys, consistent with the expected TSZ signature at WMAP
frequencies.

A more efficient detection of the SZ effect is possible by
examining the WMAP data at the location of X-ray detected clusters
since both the SZ and the X-ray data probe the same hot ICM.  In
this letter, we use the largest to-date all-sky cluster catalog,
containing  782 clusters with well measured X-ray parameters from
ROSAT All-Sky Survey data (RASS, Voges et al 1999), to determine
the TSZ amplitude present in the WMAP 3yr data and to
evaluate the properties of the cluster SZ signal. We describe in
Sec.~2 the catalog and WMAP data, present our results in Sec.~3,
and discuss their implications in Sec.~4.

\section{X-ray catalog and CMB data}

Our analysis uses an all-sky cluster sample created by combining the
ROSAT-ESO Flux Limited X-ray catalog (REFLEX, B\"ohringer et al 2004)
in the southern hemisphere, the extended Brightest Cluster Sample
(eBCS, Ebeling et al 1998, Ebeling et al 2000) in the north, and the
Clusters in the Zone of Avoidance (CIZA, Ebeling, Mullis \& Tully
2002, Kocevski et al 2007) sample along the Galactic plane.  All three
surveys are X-ray selected and X-ray flux limited using RASS data. A
detailed description of the creation of the merged catalog is provided
by Kocevski \& Ebeling (2006).

For each cluster, the catalog lists position, flux, and luminosity
measured directly from RASS data; X-ray electron temperature,
derived from the $L_X-T_X$ relation of White, Jones \& Forman
(1997), redshifts, and angular and physical extent of the region
containing the measured X-ray flux. 
We determine the X-ray extent of each cluster directly from the
RASS imaging data using a growth-curve analysis. The cumulative
profile of the net count rate is constructed for each system by
measuring the counts in successively larger circular apertures
centered on the X-ray emission and subtracting an appropriately
scaled X-ray background.  The latter is determined in an annulus
from 2 to 3 $h_{\rm 50}^{-1}$ Mpc around the cluster centroid. The
extent of each system is then defined as the radius at which the
increase in the source signal is less than the $1\sigma$
Poissonian noise of the net count rate.

To obtain an analytic parametrization of the spatial profile of
the X-ray emitting gas and, ultimately, the central electron
density we fit a $\beta$ model (Cavaliere \& Fusco-Femiano 1976)
convolved with the RASS point-spread function to the RASS data for
each cluster in our sample: $S(r) = S_0 \left[
1+(r/r_c)^2\right]^{-3\beta+1/2}$ where $S(r)$ is the projected
surface-brightness distribution and $S_0$, $r_c$, and $\beta$ are
the central surface brightness, the core radius, and the $\beta$
parameter characterising the profile at large radii. Using the
results from this model fit to determine the gas-density profile
assumes the gas to be isothermal and spherically symmetric.  In
practice, additional uncertainties are introduced by the
correlation between $r_c$ and $\beta$ which makes the results for
both parameters sensitive to the choice of radius over which the
model is fit, and the fact that for all but the most nearby
clusters the angular resolution of the RASS allows only a very
poor sampling of the surface-brightness profile (at $z>0.2$ the
X-ray signal from a typical cluster is only detected in perhaps a
dozen RASS image pixels). In recognition of these limitations, we
hold $\beta$ fixed at the canonical value of $2/3$ (Jones \&
Forman, 1984). The resulting values for $r_c$ are reassuringly
robust in the sense that we find broad agreement with the
empirical relationship between X-ray luminosity and $r_{c}$
determined by Reiprich \& B\"ohringer (1999). Our best-fit
parameters, the cluster luminosity and electron temperature, are
used to determine central electron densities for each cluster
using Eq. (6) of Henry \& Henriksen (1986) with the ICM
temperature estimated from the $L_{\rm X} - T_{\rm X}$
relationship (White et al 1997).  Other sources of error of this
parametrization, namely the deviation from a $\beta$ profile at
small radii due to cooling cores as well as the steepening of the
profile often observed at large radii (Vikhlinin, Forman \& Jones,
1999) are unlikely to affect our results due to the poor
resolution and low signal-to-noise of the RASS data. Conversions
between angular extents and the physical dimensions of clusters
are made using the $\Lambda$CDM concordance cosmology
($\Omega_\Lambda=0.7$, $h=0.7$).

In our analysis of the CMB data, we use maps from eight
differencing assemblies (DA) corresponding to the Q, V and W bands
of the ``foreground-cleaned" WMAP 3-year data: Q1, Q2, V1, V2,
W1,.., W4 available from the Lambda archive
(http://lambda.gsfc.nasa.gov). We do not use the K and Ka bands
since foreground contamination is important at those frequencies
(Bennett et al 2003). The maps are written in the HEALPix (Gorski
et al 2005) nested format with resolution $N_{side}=512$,
corresponding to pixels of $7^\prime$ on the side, significantly
larger than the $0.75^\prime$ pixels of the X-ray data, which
makes our results insensitive to deviations from spherical
symmetry of the clusters in our sample. These bands correspond to
frequencies 41, 61, and 94 GHz and angular resolutions
$\theta_{FWHM}\simeq 0.5$, 0.3, and 0.2 degrees. All maps are
multiplied by the Kp0 mask to remove microwave emission from the
Galactic plane and foreground sources. The masking eliminates 120
clusters, leaving 674 clusters for our SZ analysis. Of those, 13
clusters did not have sufficient S/N in the ROSAT data to
to define an extent and were excluded from this
analysis leaving a total of 661 clusters. 

\section{Results.}

To compute the TSZ signal from the WMAP 3-year data we evaluate
the mean temperature anisotropy at the cluster locations.  The TSZ
distortion scales as the electron density, $n_e$, integrated along
the line-of-sight, whereas the cluster X-ray emission is $\propto
n_e^2$. Thus, the clusters' SZ signal should extend over an area
significantly larger than the region within which X-ray emission
is detectable. We probe the SZ extent by measuring the signal in
regions of increasing radius $\theta_{\rm SZ}$, from 1 to
$6\theta_{\rm X}$, where $\theta_{\rm X}$ is the angular extent of
the respective cluster in the RASS data. The results from the
individual DA's were averaged with weights inversely proportional
to the pixel noise and the frequency-dependent amplitude of the SZ
effect, and without weights. The difference between the results
obtained with the different weighting schemes are smaller than 5\%
and we quote results obtained using noise and frequency-weighted
averages. Error bars are calculated assigning 1,000 random
pseudo-cluster positions to the WMAP sky and computing the mean
temperature anisotropy for each cluster template. The random
positions are always placed outside the Kp0 mask and away from any
of the cluster pixels.

In Table \ref{table1} we present our results for two subsamples:
clusters with redshift $z\le 0.2$ and clusters with luminosity
$L_X([0.1-2.4]$keV$)\ge 3\times 10^{44}$erg/s. The limit of $z$ =
0.2 was chosen since this is the largest redshift out to which our
cluster catalog is reasonably complete. We use the average values
of the core radii and radial cluster extent for each subsample.
The TSZ signal is measured within (a) a disc of angular radius
$\theta$, and (b) a ring or annulus delimited by two consecutive
discs, to differentiate the contribution coming from regions
increasingly farther away from the cluster core.  By design, the
first DISC and RING regions are identical. We measure a
temperature decrement at the cluster pixels at the $\sim 15\sigma$
level for the most X-ray luminous clusters. The region from 3-4
times the X-ray extent still shows a non-negligible contribution,
while outside this region no statistically significant signal is
detected.  The average radius of the regions contributing to the
signal is $\simeq 2-3h_{70}^{-1}$Mpc, depending on the cluster
sample. The fact that clusters appear more extended at microwave than at X-ray
frequencies was used by us (Kashlinsky et al 2008) to isolate the KSZ
component arising from the cluster bulk motion; as we demonstrate below
the cluster X-ray temperature decreases towards cluster outskirts making
the KSZ dipole appear still more extended than the TSZ contributions.

\begin{figure}[ht]
\plotone{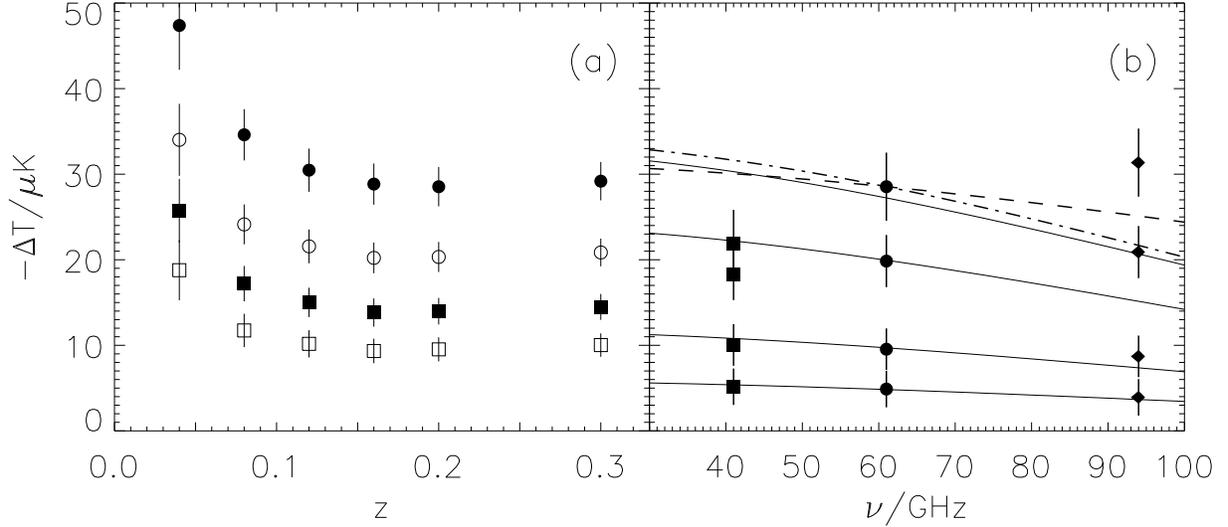}
\caption[]{ (a) TSZ signal in WMAP 3-year data for cluster subsamples
  with redshift $z\le [0.04,0.08,0.12,0.16,0.2,0.3]$. Close and open
  circles, squares and diamonds correspond to clusters with different
  angular extent. From top to bottom, the average was taken from
  clusters with total extent $\theta_{\rm SZ}=(1-4)\times\theta_{\rm X}$.
  (b) TSZ emission vs frequency is shown for different $\theta_{\rm SZ}$ for: Q
  (squares), V (circles)
  and W (diamonds) bands and clusters with $z\le 0.2$.  Solid lines
  show the frequency dependence of the TSZ effect.
  From top to bottom $\theta_{\rm SZ}=(1,2,4,6)\times\theta_{\rm X}$.
  In all cases, vertical bars indicate the $1\sigma$ errors.
  Curves were normalized by least squares regression.
  Dashed and dot-dashed lines show the
  frequency dependence of the KSZ and Relativistic SZ effects, respectively,
  normalized to the amplitude measured at the V-band.
} \label{fig1}
\end{figure}

In Fig.~\ref{fig1} we show the TSZ amplitude for different cluster
subsamples and regions of different radius. In Fig.~\ref{fig1}a,
we plot the results for clusters within a progressively increasing
upper redshift limit; the last data point includes all clusters.
Except for the lowest redshift bins the measured amplitude is
roughly independent of the number of clusters. In Fig.~\ref{fig1}b
we show the frequency dependence of the measured TSZ signal for
all clusters with redshift $z\le 0.2$ by comparing the result for
the three bands considered: Q, V, and W. Solid lines denote the
frequency dependence of the TSZ effect with the curves normalized
by least squares regression. Note that for the smallest angular
extent the TSZ signal shows frequency spectrum which differs
significantly from the TSZ expectation. Earlier analysis by
Hern\'andez-Monteagudo et al (2004) also did not find a clear TSZ
scaling in WMAP 1st-year data. This difference could originate due
to the different angular resolution and noise levels between the
Q, V and W bands. Due to its smaller resolution, the Q channel
probes the SZ decrement to a larger extent and dilutes the central
SZ amplitude more than in the V or W bands. 
As we include the
contribution from progressively larger radii 
the effect of noise and resolution is reduced;
by the time we reach $\theta_{\rm SZ}= 4-6\theta_{\rm
X}$, the change in frequency at the 3 bands agrees very well with
the spectral dependence of the TSZ effect. 
Other components have negligible contribution.
A bulk flow motion will give rise to a dipole
pattern but does not change the monopole evaluated at cluster
locations. In the Rayleigh-Jeans regime, the first correction to
the TSZ frequency scaling due to the relativistic effect is
$1.7(k_BT_X/m_ec^2)$ (see Itoh, Kohyama \& Nozawa, 1998). The
average temperature of all clusters in our catalog is $k_BT_X\sim
4$KeV, so the relativistic correction represents less than a 1.4\%
variation. Dashed and dot-dashed lines in Fig.~\ref{fig1}b
represent the frequency dependence of the KSZ and relativistic SZ
effect, the latter computed using the approximate solution of Itoh
et al (1998) for a cluster of $k_B T_X=10$KeV, arbitrarily
normalized to the amplitude of the V-band.

\begin{figure}[ht]
\plotone{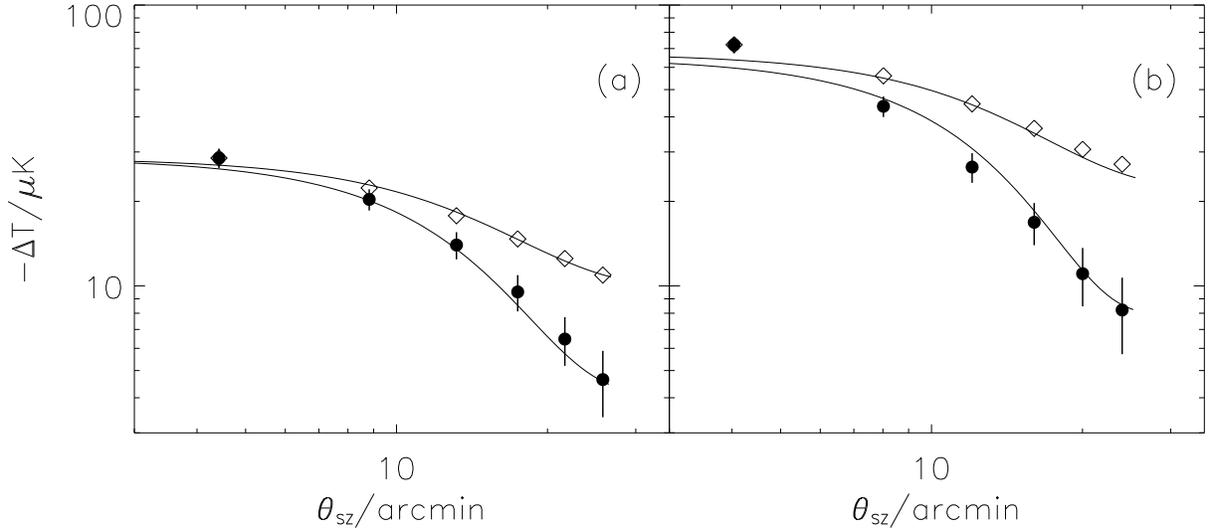}
\caption[]{Measured and predicted electron pressure profile vs angular
  cluster-centric radius. Full circles and open diamonds correspond to
  the measured profile and the profile predicted for isothermal
  $\beta=2/3$ model, respectively. Solid lines correspond to the $\beta$
  model (upper) and the NFW model fit (lower) assuming a single cluster
  at $z=0.12$. We show results for two cluster subsamples:
  (a) $z\le 0.2$, (b) $L_x[0.1-2.4KeV]\ge 3\times 10^{44}$ ergs$^{-1}$.
  In (a) the best fit corresponds to $c=8$ and $r_s=350h^{-1}$kpc
 and in (b) to $c=15$ and $r_s=250h^{-1}$kpc.} \label{fig2}
\end{figure}

The radial emission profiles derived from SZ and X-ray
observations of clusters do not necessarily follow the same
$\beta$ model. Only a handful of clusters have measured radial SZ
profiles and the parameters obtained from X-ray data are often
used in SZ analysis. Using numerical simulations, Hallman et al
(2007) showed that this approach leads to biased estimates of the
integrated Compton y-parameter in the inner part of clusters.
Since we measure the TSZ contribution outside the inner cluster
region, we can test for a similar bias in the outskirts of
clusters and assess the accuracy of the $\beta$ model. In
Fig.~\ref{fig2} a comparison is made between our measured radial
SZ profile (full circles, see also Table \ref{table1}) and $\beta$
model predictions (diamonds). Predictions are computed with the
same pipeline as the data from the eight DA maps generated by
placing clusters on the sky with their measured angular size. To
each pixel within a cluster we assign CMB temperature derived from
the $\beta$ model TSZ profile, convolved with the WMAP beam for
each DA. The angular scale $\theta_{\rm SZ}$ is the area weighted
average extent of cluster.

The discrepancy between data and the $\beta=2/3$ prediction is less than a few
$\mu K$ or $10-30\%$ in the cluster centers, but increases with
radius and reaches a factor of $2-3$ at the largest radius probed
in our study. It could be argued that a model
with $\beta=1$ would fit the data in the cluster outskirts.
Such a high $\beta$ value would not only invalidate the
values of $r_c$ and $n_c$ derived directly from the RASS X-ray
data, it would in fact be inconsistent with the shape of the X-ray
surface brightness profile in the central region of essentially
every cluster ever observed. 
Jones and Forman (1999) found that the average cluster surface-brightness
profile is well described by a $\beta$ model with $r_c\sim 200$Kpc
and $\beta\sim 0.6$. Out of 96 clusters analyzed in detail, they
did not find a single one for which a $\beta$ value outside the
range of 0.4--0.8 would provide an acceptable fit to the X-ray data.

Numerical hydrodynamic simulations suggest that the dark-matter
distribution in galaxy clusters is described by a universal
density profile (Navarro, Frenk \& White 1997, NFW),
$\rho_{dm}(x)=\rho_s/[x(1+x)^2]$, where $x=r/r_s$, $r_s$ and
$\rho_s$ are a characteristic scale radius and density. Usually
$r_s$ is given in terms of the concentration parameter $c=r_v/r_s$
where $r_v$ is the halo virial radius. This parameter depends only
weakly on mass, with less massive systems being more concentrated,
having larger $c$. While the electron density for the $\beta=2/3$
model scales as $r^{-2}$ at large radii, the NFW is much steeper,
scaling as $r^{-3}$.  If the gas distribution were to follow that
of the dark matter, one would thus expect its radial profile to
decline much more steeply, as observed and shown in
Fig.~\ref{fig2}. Solid lines represent the electron pressure
profile of a single cluster computed using a $\beta$ (upper) and a
NFW model (lower solid line), convolved with the WMAP 3-year beam.
Our fits ($\beta$ and NFW model alike) assume a representative
cluster redshift of $z=0.12$, a value close to the mean redshift
of both cluster subsamples. The best-fit values for the core
radius of the $\beta$ model are $\theta_c=1.5',0.5'$ (a and b,
respectively). We checked that no value of $r_c\in[0.5,5]$arcmin
can make the $\beta$ model fit the measured pressure profile. To
generate the NFW profile, we follow Komatsu \& Seljak (2001, 2002)
and assume that the gas follows the DM distribution, is in
hydrostatic equilibrium, and is well described by a polytropic
equation of state. Fitting an NFW model yields best-fit
concentration parameters of $c=8$ (a) and $c=15$ (b). In the
Komatsu \& Seljak (2001) model, these values correspond to
polytropic indices $\gamma=1.17,1.2$, respectively. This result
reflects that the most luminous clusters are, on average, further
away in a flux limited sample than the whole population, so they
subtend a smaller angular size and appear concentrated (smaller
$r_s$ in NFW) or have a smaller core radius (in the $\beta$
model). In Fig.~\ref{fig2}, theoretical lines represent the TSZ
signal of a single cluster; as such, they are not a fair
representation of the cluster population as a whole. If
$\theta_{\rm SZ}$ were not the area weighted extent but the SZ
emmision weighted center, the best fit NFW and $\beta$ models
-solid lines in the figure- would correspond to slightly different
model parameters but it will not change the discrepancy between
the measured and the $\beta$ model predicted profiles.

The universal gas temperature profiles of Komatsu \& Seljak
(2001), compatible with our results, show a strong decline of gas
X-ray temperature with radius. The central temperature decreases
by a factor 2-4 at the virial radius, being steeper for the more
concentrated (less massive) clusters. This result is in agreement
with the recent analysis of X-ray temperature profiles of 15
nearby clusters, carried out by Pratt et al (2007) using
XMM-Newton data. They measured temperature profiles declining by a
factor $\sim 2$ at half the virial radius, in good agreement with
numerical simulations outside the core region. To conclude,
Fig.~\ref{fig2} shows that the gas density has a steeper
decline in the outer region than an isothermal $\beta=2/3$ model;
the slope is close to -3, i.e., the dynamical
state of cluster outskirts is well described by a NFW profile.

\section{Conclusions.}

Using the largest X-ray cluster catalog available today we present
an accurate measurement of the contribution of clusters of
galaxies to the temperature anisotropies measured by the WMAP
satellite in its three-year data release. We find the TSZ signal
to extend to, on average, $\sim 2-3h_{70}^{-1}$Mpc, i.e., radii
much larger than the ones out to which X-ray emission is
detectable. The TSZ signal measured in the cluster cores shows
deviations the expected TSZ frequency behavior. These are likely
caused by the different angular resolution of the WMAP Q, V, W
channels resulting in probing SZ decrement from different parts of
the clusters. However, when the TSZ contribution from the outer
cluster regions is included, the signal is consistent with the
frequency dependence of a TSZ spectrum.

The measurement indicate that the gas profile of the cluster
population in the outer region is compatible with the NFW model.
Our results suggests than the cluster temperature profile declines
with radius, in agreement with numerical simulations of clusters
and recent XMM data.  We are currently rederiving the NFW data
parameters from the X-ray images to compare the TSZ predicted for
the cluster population using X-ray derived quantities with the
signal at WMAP frequencies. The radial profile of the measured TSZ
signal suggests that, all the way to the cluster outskirts,
baryons are settled in hydrostatic equilibrium within the DM
potential well and follow the same density distribution. As shown
by numerical simulations, the profile of collapsed dark matter
haloes is a direct consequence of the Newtonian gravity with a
suitably chosen initial density field (corresponding to the
concordance $\Lambda$CDM model). Hence our results also provide
further, albeit indirect, evidence for the existence of dark
matter and the validity of the Newtonian dynamics.

\acknowledgments

This work is supported by the Ministerio de Educaci\'on y Ciencia
and the ''Junta de Castilla y Le\'on'' in Spain (FIS2006-05319,
PR2005-0359 and SA010C05) and by NASA ADP grant NNG04G089G. We
thank Gary Hinshaw for useful information regarding the WMAP data
specifics. FAB thanks the University of Pennsylvania for its
hospitality when part of this work was carried out.

\clearpage
\begin{table}
\begin{tabular}{|cc|rr|cc|rr|}
\hline
\multicolumn{4}{|c|}{$z_{cl}\le 0.2$}&\multicolumn{4}{c|}{$L_X\ge 3\times 10^{44}erg/s$} \\
\hline
\multicolumn{2}{|c|}{Data}&\multicolumn{2}{c|}{TSZ (in $\mu$K)}&
\multicolumn{2}{|c|}{Data}&\multicolumn{2}{c|}{TSZ (in $\mu$K)} \\
$\theta$/arcmin&$r_{ext}/$Mpc&Disc\hspace*{.5cm}&Ring\hspace*{.5cm}&
$\theta$/arcmin&$r_{ext}/$Mpc&Disc\hspace*{.5cm}&Ring\hspace*{.5cm}\\
(1) & (2) & (3) & (4) & (5) & (6) & (7) & (8) \\
 \hline
   4.4 & 0.6 & -28.5 $\pm$ 2.3 &  -28.5 $\pm$  2.3 &
   4.0 & 1.0 & -72.2 $\pm$ 4.8 &  -72.2 $\pm$  4.8 \\
   8.8 & 1.3 & -20.3 $\pm$ 1.8 &  -17.4 $\pm$  2.0 &
   8.0 & 1.9 & -43.6 $\pm$ 3.7 &  -33.5 $\pm$  4.2 \\
  13.2 & 1.9 & -14.0 $\pm$ 1.6 &   -8.4 $\pm$  2.1 &
  12.0 & 2.9 & -26.5 $\pm$ 3.2 &  -12.8 $\pm$  4.3 \\
  17.4 & 2.5 &  -9.5 $\pm$ 1.4 &   -3.1 $\pm$  2.1 &
  16.0 & 3.8 & -16.9 $\pm$ 2.9 &   -4.4 $\pm$  4.4 \\
  21.7 & 3.1 &  -6.5 $\pm$ 1.3 &   -0.5 $\pm$  2.1 &
  20.0 & 4.8 & -11.1 $\pm$ 2.6 &   -0.8 $\pm$  4.3 \\
  25.8 & 3.8 &  -4.6 $\pm$ 1.2 &    0.4 $\pm$  2.2 &
  24.0 & 5.7 &  -8.2 $\pm$ 2.5 &   -1.7 $\pm$  4.5 \\
\hline
\end{tabular}
\caption{ TSZ amplitude evaluated at cluster locations for two cluster
  samples. Columns: (1,2,5,6) show the average angular and physical cluster
  radius for each subsample; (3,4,7,8) correspond to the measured TSZ signal in the
  WMAP-3yr data at the cluster locations.  In DISC we measure the TSZ
  signal within a circle of angular radius $\theta_{\rm
    SZ}=[1-6]\theta_{\rm X}$, in RING within an annulus of width
  $1\theta_{\rm X}$ and increasing radius.}
\label{table1}
\end{table}

\clearpage

\clearpage

\end{document}